\documentstyle[psfig]{mn}
\title[Brown dwarfs de-frocked]
{Optical spectroscopy of a brown dwarf candidate
\thanks{Based partially on observations made with the Keck telescope}}
\author[I. N. Reid ]
{I. Neill Reid $^{1}$ \\
$^{1}$ Palomar Observatory, 105-24, Caltech, Pasadena, CA 91125, USA\\
email: inr@dowland.caltech.edu\\}

\begin{document}

\maketitle

\begin{abstract}

We have used the Low-Resolution Imaging Spectrograph on the 
Keck II telescope to observe the brown dwarf candidate D04
(Hawkins et al, 1998). The spectrum matches that of a 
spectral-type M7 dwarf, implying a photospheric temperature of
$\approx 2600$K. This is consistent with the available (R-I)$_C$
and (I-K) colours. If the parallax measured by Hawkins et al is
correct, then the implication is that D04 has a radius of $\sim 0.035 R_\odot$,
or one-third that of Jupiter. This contradicts the predictions made by
current stellar models that electron degeneracy leads to nearly
constant radii for stars and brown dwarfs at masses below 0.1 M$_\odot$.
We suggest that an equally valid interpretation of the data is that
D04 is a VB8 analogue at a distance of $\approx 150$ parsecs.

\end{abstract}


\section{introduction}

While brown dwarfs do not supply the missing mass, their properties continue to be 
of considerable importance for our understanding of star formation and
stellar evolution. the discovery of Gl 229B (Nakajima et al, 1995) and of
a variety of sub-stellar mass objects in the Pleiades (Rebolo et al, 1995)
confirmed the existence of these objects, but, save for Kelu 1 (Ruiz et al, 1997),
isolated objects in the field remained elusive. Identifying these intrinsically faint,
cool objects requires deep, wide-field imaging at red or near-infrared wavelengths,
as emphasised by the initial results from $1-2 \mu m$ DENIS (Delfosse et al, 1997) and
2MASS (Kirkpatrick et al, 1998) surveys. \\

Prior to the availability of large-scale near-infrared photometry, photographic plates
offered the only viable method of surveying tens or hundreds of square degrees. Such media
are limited to wavelengths shortward of 1$\mu m$, but can achieve (single-plate)
detection limits of R$_C \sim$ 21 to 21.5 and I$_C \sim 19$ to 19.5 magnitudes. 
Moreover, Hawkins has experimented with digital addition of plate scans, and finds
that the limiting magnitude can be extended significantly ($>2$ magnitudes) if
20-40 plates are available for a given field. Although expensive in telescope
time, this technique makes photography competitive with optical or
near-infrared CCD imaging for a number of specific projects. \\

Hawkins has concentrated analysis on a single field, ESO/SERC field 287 centred at
$\alpha = 21^h 28^m$, $\delta = -45^o$. Recently, Hawkins et al (1998) reported on
initial results from searching a combination of 65 IIIaF and 30 IVN plates for
candidate very low mass stars or brown dwarfs. They announce the discovery of
at least three brown dwarfs with somewhat unusual properties: the optical and
near-infrared colours match those of late-type M-dwarfs, but the absolute magnitudes,
calibrated using CCD-derived trigonometric parallaxes, place the objects $\sim 2.5$
magnitudes below the main-sequence. In contrast, theoretical evolutionary calculations
(e.g. Burrows et al, 1997) predict that cooling brown dwarfs should lie {\sl above}
the stellar main-sequence at these temperatures. Hawkins et al suggest that their
candidates may be either metal-poor or subject to unusual dust formation in the
atmosphere. \\

We present here optical spectroscopy of one of the three brown dwarf candidates, D04. 
The following section describes our observations, while the final section summarises 
our conclusions.

\begin{figure}
\psfig{figure=fig1.ps,height=10cm,width=8cm
,bbllx=8mm,bblly=57mm,bburx=205mm,bbury=245mm,rheight=12.0cm}
\caption{An i-band finding chart for the 2 arcminute field centred on D04}
\end{figure}

\begin{figure}
\psfig{figure=fig2.ps,height=11cm,width=8cm
,bbllx=8mm,bblly=57mm,bburx=205mm,bbury=245mm,rheight=13.0cm}
\caption{ The far-red optical spectrum of D04}
\end{figure}

\section{Spectroscopic Observations}

Our observations of D04 were obtained on August 11, 1998 as part of a service
allocation with the Low-Resolution Imaging Spectrograph (Oke et al, 1995) 
on the Keck II telescope. D04 (R $\sim 22$, I$\sim 19.3$) was not directly visible on
the acquisition TV. However, we used the imaging capability of LRIS to
obtain a 60-second I-band frame which, combined with Hawkins et al's position and
the Digital Sky Survey scan, allowed unambiguous identification of the target
(figure 1). We then offset the telescope from a brighter star, placing D04 on 
the 1-arcsecond slit for spectroscopic observation. \\

Spectroscopy was undertaken using a 400 l/mm grating, blazed at 8500\AA,
with a central wavelength of 8000 \AA. This provides coverage from $\lambda \sim 6200 \AA$\
to $\sim 9800 \AA$\ at a dispersion of 1.85 \AA\ pix$^{-1}$ and a resolution of
4.5 pixels. Wavelength calibration is provided by neon-argon arclamp exposures, taken
immediately after the stellar integrations. The data were bias-subtracted and
flatfielded, and the spectra extracted using standard IRAF software. \\

We obtained a single 1800-second exposure of D04, which we flux-calibrated 
using an observation of the white dwarf standard G24-9 (Filippenko \& Greenstein, 1983). 
Seeing was $\sim1$ arcsecond, even at an altitude of 25$^o$. However, the slit was
not aligned with the parallactic angle, and the overall shape of the D04 spectrum is
likely to be affected by differential chromatic refraction ($\sim0.6$ arcseconds between 
6500 and 9500\AA\ at sec(z)=2.3). Despite the
faint apparent magnitude and the low altitude of the target, the extracted 
spectrum has a signal-to-noise of at least 15 for $\lambda > 7300 \AA$. 

\section {Discussion and conclusions}

Figure 2 plots the flux-calibrated spectrum of D04. The object is clearly of
spectral type M, with TiO bandheads at $\lambda\lambda 7666, 8432$ and  8859 \AA, 
as well as strong absorption (EW $\sim 11.6$\AA) due to the sodium doublet at $\lambda 8183/8195$\AA. 
Hawkins et al
suggested that the object might be metal-poor and/or subject to unusual dust 
obscuration. However, the strength of the TiO absorption rules out the 
possibility that D04 is a late-type subdwarf, comparable to LHS 1742a (Gizis, 1997).
Similarly, a comparison with the spectral standard M-dwarfs  defined by
Kirkpatrick et al (1991, 1993) shows no evidence for unusual absorption, which might
be attributed to excessive dust formation.
The presence of significant VO absorption at $\lambda\lambda 7334$ and 7850 \AA\ indicates 
that the spectral type is later than M5, and a visual comparison with the
Kirkpatrick et al standard sequence leads to classification as between M7 and
M8. A spectrum of the well-known M8 dwarf VB10 is plotted in figure 2 for comparison. 
Extrapolating the results derived by Leggett et al (1996), d04 is likely to have
T$_{eff} \approx 2600$K. \\

VB8 (Gl 644C) is the best-calibrated M7 standard in the Kirkpatrick et al system. 
Allowing for the uncertainties in Hawkins et al photographic photometry, D04 
has an (R-I)$_C$ colour (2.63 mag.) consistent with that of VB8 (2.41 mag. - Leggett, 1992), 
while the (I-K) colours of the two stars are nearly identical at 3.7 magnitudes. However,
VB8 has an absolute magnitude at 2.2$\mu m$ of M$_K = 9.76$, while Hawkins
et al deduce M$_K = 12.24$ for D04. Given the similarity in spectral types
and optical/IR colours, as well as the absence of evidence for any chemical 
peculiarities in D04, it is reasonable to assume that the two objects have similar
effective temperatures and similar bolometric corrections. In that case,

$$ \Delta L \qquad \propto \quad \Delta R^2 $$

That is, the difference in luminosity of 2.5 magnitudes inferred by Hawkins et al
implies that D04 has a radius which is three times smaller than that of VB8. \\

Leggett et al (1996) have combined optical and infrared spectroscopy and
photometry with improved model atmospheres to derive effective temperatures,
luminosities and radii for a small number of M dwarfs. The lowest luminosity (and
lowest temperature) star in their sample is GJ 1111, spectral type M6.5, M$_K \sim 9.46$, for which
they estimate a radius of $0.8 \times 10^8$ metres. this corresponds to 0.11 R$_\odot$, or slightly 
less than the radius of Jupiter (0.119 R$_\odot$). Assuming a similar radius 
for the slightly later-type VB8, the luminosity deduced by Hawkins et al for D04
leads us to infer a radius of $\sim 0.035 R_\odot$, or one-third that of Jupiter. \\

This result is clearly at odds with predictions based on interior models of low-mass 
stars and brown dwarfs. As the mass decreases towards 0.1 M$_\odot$, theoretical models
predict that the radius also decreases to close to 0.12 R$_\odot$ (Burrows \& Liebert, 
1993, figure 1). However, electron degeneracy takes over as the main source of
pressure support in lower-mass objects, and, as a result, the radius is
predicted to vary by no more than $\sim 30\%$ as the mass decreases to one 
Jupiter mass (Burrows et al, 1997). Moreover, substellar-mass objects ($M < 0.075 M_\odot$)
are predicted to have radii {\sl exceeding} that of Jupiter at effective
temperatures of 2600K. \\

Given that there is no evidence for unusual atmospheric opacities in D04, and
that the deduced radius is in strong contradiction with a basic premise of
stellar structure, an alternative explanation must be found for 
the faint absolute magnitudes deduced by Hawkins et al. The simplest is
that the trigonometric parallax derived by Hawkins et al for at least D04
(and possibly D07 and D12) overestimates the true value. Each star was
observed at only three epochs, leading to astrometric solutions which
are poorly constrained against systematic errors (cf. Pinsonneault et al's 
(1998) comments on the Hipparcos Pleiades astrometry). Moreover, there is
significant dispersion amongst the individual astrometric measurements at
a given epoch for each of the three faint (I$\sim 19.4$) candidate brown
dwarfs (Hawkins et al, figure 6). Finally, Tinney (priv. comm.) points out that
the differential chromatic refraction corrections are constrained poorly,
raising the possibility of systematic errors in the final astrometric 
solution. \\

Further astrometry of these objects is clearly desirable, but for the present, 
we favour interpreting the current data in alternative manner to the solution
espoused by Hawkins et al. We identify D04, D07 and D12
as M7/M8 main-sequence disk dwarfs, lying at distances of $\sim 150$ parsecs,
rather than as highly-unusual brown dwarfs at distances of $\sim 50$ parsecs.

\subsection*{Acknowledgements}

I would like to thank Greg Wirth and Gary Puniwai for assistance with 
the observations. The Keck Observatory
is operated by the Californian Association for Research in Astronomy, and
was made possible by generous grants from the Keck W. M. Foundation.


\begin{thebibliography}{}
\bibitem[]{} Burrows, A., Liebert, J. 1993, Rev. Mod. Phys. 65, 301
\bibitem[]{} Burrows, A., Marley, M., Hubbard, W.B., Lunine, J.I., Guillot, T., Saumon, D.,
Freedman, R., Sudarsky, D., Sharp, C. 1997, ApJ, 491, 856
\bibitem[]{} Delfosse, X., Tinnet, C.G., Forveille, T., Epchtein, N., Bertin, E. et al 1997, AA, 327, L25
\bibitem[]{} Filippenko, A.V., Greenstein, J.L. 1983, PASP, 96, 530
\bibitem[]{} Gizis, J.E. 1997, AJ, 113, 806
\bibitem[]{} Hawkins, M.R.S., Ducourant, C., Jones, H.R.A., Rapaport, M. 1998, MNRAS, 294, 505
\bibitem[]{} Kirkpatrick, J.D., Henry, T.J., McCarthy, D.W.Jr. 1991, ApJS, 77, 417
\bibitem[]{} Kirkpatrick, J.D., Kelly, D.M., Rieke, G.H., Liebert, J., Allard,  Wehrse, R. 
1993, ApJ, 402, 643
\bibitem[]{} Kirkpatrick, J.D., Reid, I.N., Liebert, J.L., Curti, C., Beichmann, C., .. 1998,
ApJ, submitted
\bibitem[]{} Leggett, S.K. 1992, ApJS, 82, 351
\bibitem[]{} Leggett, S.K., Allard, F., Berriman, G., Dahn, C.C., Hauschildt, P. 1996, ApJS, 104, 117
\bibitem[]{} Nakajima, T., Oppenheimer, B.R., Kulkarni, S.R., Golimowski, D.A.,
Matthews, K., Durrance, S.T. 1995, Nature, 378, 463
\bibitem[]{} Oke, J. B., Cohen, J. G., Carr, M., Cromer, J., Dingizian, A., 
Harris, F. H., Labreque, S., Lucinio, R., Schaal, W., Epps, H., Miller, J. 1995, PASP, 107, 375
\bibitem[]{} Pinsonneault, M.H., Stauffer, J., Soderblom, D.R., King, J.R., Hanson, R.B. 1998, ApJ, in press
\bibitem[]{} Rebolo, R., Zapatero-Osorio, M.R., Martin, E.L. 1995, Nature, 377, 129
\bibitem[]{} Ruiz, M.T., Leggett, S.K., Allard, F. 1997, ApJ, 491, L107

\end{thebibliography}
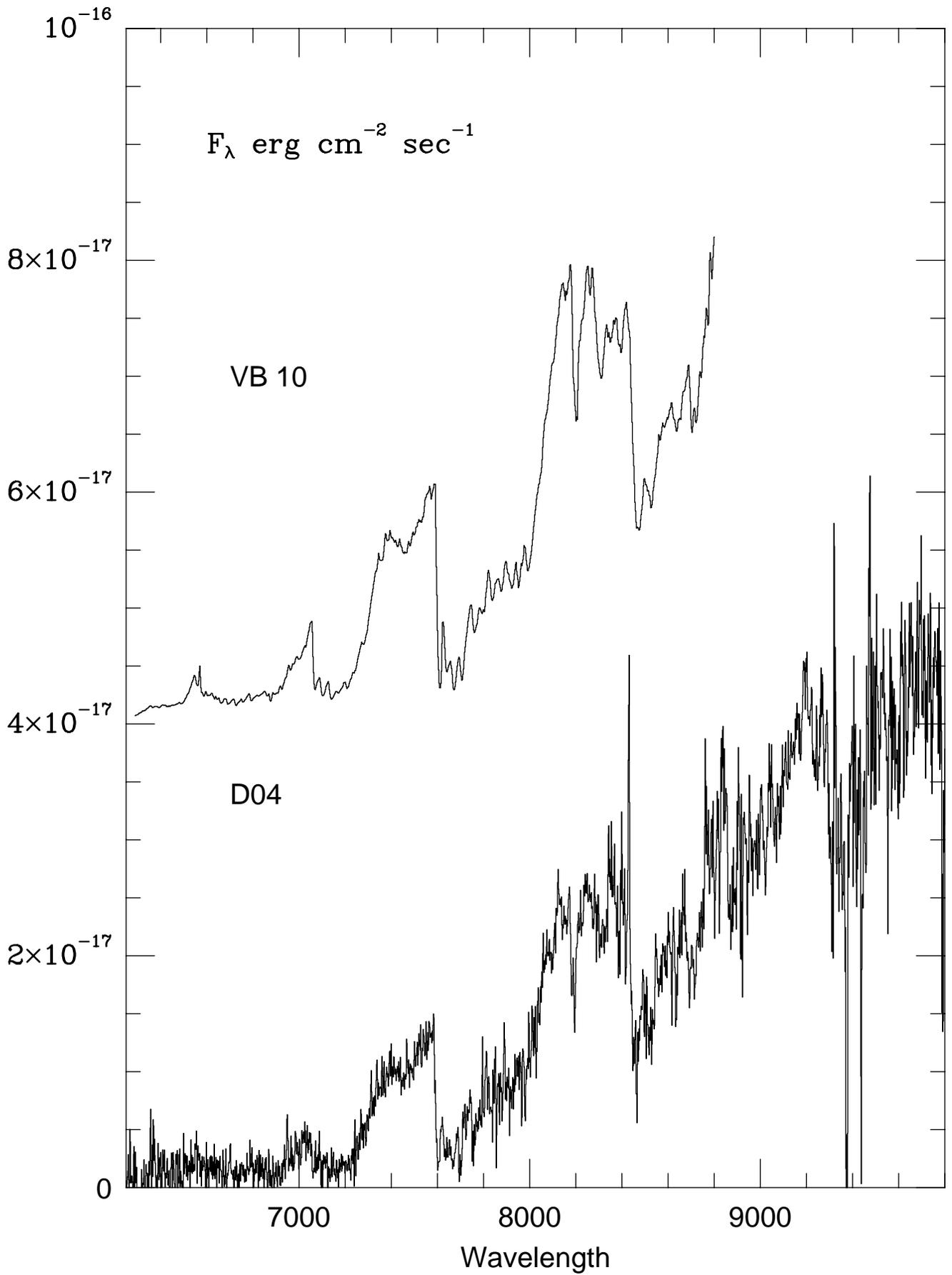
\end{document}